\documentclass[acp, manuscript]{copernicus}

\usepackage{latexsym,euscript,textcomp}
\usepackage{color,graphics,epsf,dcolumn}
\usepackage{epstopdf,ulem}

\begin{document}
\nolinenumbers

\title{Hurricane's maximum potential intensity and surface heat fluxes}

% \Author[affil]{given_name}{surname}

\Author[1,2]{Anastassia M.}{Makarieva}
\Author[1,2]{Victor G.}{Gorshkov}
\Author[1]{Andrei V.}{Nefiodov}
\Author[3]{Alexander V. }{Chikunov}
\Author[4]{Douglas}{Sheil}
\Author[5]{Antonio Donato}{Nobre}
\Author[6]{Paulo}{Nobre}
\Author[2]{Bai-Lian}{Li}

\affil[1]{Theoretical Physics Division, Petersburg Nuclear Physics Institute, Gatchina  188300, St.~Petersburg, Russia}
\affil[2]{USDA-China MOST Joint Research Center for AgroEcology and Sustainability, University of California, Riverside 92521-0124, USA}
\affil[3]{Princeton Institute of Life Sciences, Princeton, New Jersey 08540, USA}
\affil[4]{Faculty of Environmental Sciences and Natural Resource Management, Norwegian University of Life Sciences, \AA s, Norway}
\affil[5]{Centro de Ci\^{e}ncia do Sistema Terrestre INPE, S\~{a}o Jos\'{e} dos Campos, S\~{a}o Paulo  12227-010, Brazil}
\affil[6]{Center for Weather Forecast and Climate Studies INPE, S\~{a}o Jos\'{e} dos Campos, S\~{a}o Paulo 12227-010, Brazil}

%% The [] brackets identify the author with the corresponding affiliation. 1, 2, 3, etc. should be inserted.

\runningtitle{Hurricane's maximum potential intensity and surface heat fluxes}

\runningauthor{Makarieva et al.}

\correspondence{A. M. Makarieva (ammakarieva@gmail.com)}

\received{}
\pubdiscuss{} %% only important for two-stage journals
\revised{}
\accepted{}
\published{}

%% These dates will be inserted by Copernicus Publications during the typesetting process.

\firstpage{1}

\maketitle

\section*{Key Points:}
\begin{itemize}
\item  Neglecting kinetic energy in the outflow results in Emanuel's Potential Intensity, here re-derived, underpredicting storm's maximum velocity
\item  A revised maximum velocity estimate is shown to depend on oceanic latent heat flux only, independent of sensible heat or dissipative heating
\item  Contrary to previous research, the energy needed to lift precipitating water is shown to have little impact on storm intensity
\end{itemize}

\begin{abstract}
Emanuel's concept of Maximum Potential Intensity (E-PI) relates the maximum velocity $V_{\rm max}$ of tropical storms, assumed to be in gradient wind balance, to environmental parameters. Several studies suggested that the unbalanced flow is responsible for E-PI sometimes significantly underpredicting $V_{\rm max}$.  Additionally, two major modifications generated a considerable range of E-PI predictions:  the dissipative heating and the power expended to lift water were respectively suggested to increase and reduce E-PI $V_{\rm max}$  by about 20\%.   Here we re-derive the E-PI concept separating its dynamic and thermodynamic assumptions  and lifting the gradient wind balance limitation.
Our analysis reveals that E-PI formulations for a balanced and a radially unbalanced  flow are similar, while the systematic underestimate of $V_{\rm max}$ reflects instead an incompatibility between several E-PI assumptions. We discuss how these assumptions can be modified. We further show that irrespective of whether dissipative heating occurs or not, E-PI uniquely relates $V_{\rm max}$ to the latent heat flux (not to the total oceanic heat flux as originally proposed). We clarify that, in contrast to previous suggestions, lifting water has little impact on E-PI. We demonstrate that in E-PI the negative work of the pressure gradient in the upper atmosphere consumes all the kinetic energy generated in the boundary layer. This key dynamic constraint is independent of other E-PI assumptions and thus can apply to diverse circulation patterns.
Finally, we show that the E-PI maximum kinetic energy per unit volume equals the local partial pressure of water vapor and discuss the implications of this finding for predicting $V_{\rm max}$.
\end{abstract}

%\Large
\introduction  %% \introduction[modified heading if necessary]

\label{intr}

Predicting hurricane intensity is a challenge. In the modern literature
{\it intensity} can refer either to the maximum pressure fall \citep[e.g.,][]{malkus60,em88,holland97,hart07}
or to the maximum sustained velocity within a storm \citep[e.g.,][]{em86,CampMontgomery01,kowaleski16}.

Early theoretical studies focused on pressure.  Given that the hurricane is warmer than the ambient environment the idea was to
retrieve the surface pressure deficit from this extra warmth assuming the existence of an unperturbed atmospheric top where pressures in the hurricane and the environment coincide. Since air pressure drops with altitude more slowly when the atmosphere is warm than when it is cold, to arrive at equal pressures at the top of the troposphere one must start from a lower surface pressure in the warmer column. The height of the unperturbed top and the extra warmth of the hurricane compared to its environment uniquely determined the  surface pressure deficit in the storm.  While this pressure deficit is well correlated  with maximum velocity   \citep{holland80,Willoughby06,knaff07,holland08,kossin15,chavas17},
it was challenging to describe this correlation from theory rather than observational fitting and thus retrieve a velocity value from the predicted pressure deficit.

\citet{em86} noted that the work associated with surface pressure deficit, and available to generate winds,
is constrained not only by the first law of thermodynamics, but also by the Bernoulli equation. Combining these and assuming that both the generation of kinetic energy and its dissipation (proportional to the cube of velocity) occur within the boundary layer, he linked work to power to obtain explicit formulae for calculating maximum hurricane velocity from environmental parameters.
This upper limit became widely used (see recent discussions by \citet{garner15}, \citet{kieu2016} and \citet[][]{kowaleski16}).

The key equation of Emanuel's potential intensity concept (E-PI hereafter) relates the local surface flux of turbulent dissipation $D$ (W~m$^{-2}$) and the oceanic heat flux $J$ (W~m$^{-2}$)
\begin{gather}
D =  \rho C_D V^3,    \label{D} \\
J =  \rho C_k V(k_s^* - k) = \rho C_k V (c_p \Delta T + L \Delta q),  \label{J}
\end{gather}
at the radius of maximum wind $V = v_{\rm max}$:
\begin{equation}\label{JD}
D = \varepsilon_C J,
\end{equation}
such that
\begin{equation}\label{vmax}
v_{\rm max}^2 = \varepsilon_C \frac{C_k}{C_D} (k_s^* - k).
\end{equation}
Here $V$ and $v$ is total  and tangential air velocity, $\varepsilon_C = (T_a-T_o)/T_a$ is the efficiency of a Carnot cycle operating on a temperature difference between the ambient temperature $T_a$ in the boundary layer and temperature $T_o$ of air outflow in the upper at\-mos\-phere; $\rho$ is air density, $C_k \approx C_D \sim 10^{-3}$ are surface exchange coefficients for enthalpy and momentum, respectively; $k_s^*$ (J~kg$^{-1}$) is saturated enthalpy of air at surface temperature and $k \approx c_p T_b + L q_b$ is the actual enthalpy of air in the boundary layer; $\Delta q \equiv q_s^* - q_b$ is the difference between saturated water vapor mixing ratio $q_s^*$ at the oceanic surface and the actual mixing ratio $q_b$ in the boundary layer; $\Delta T \equiv T_s - T_b$ is, likewise, the difference between temperature $T_s$ of the oceanic surface and temperature $T_b$ of the adjacent air in the boundary layer, $v_{\rm max}$ is maximum tangential velocity, which at the radius of maximum wind approximates total velocity (the vertical and radial velocities are assumed to be relatively small).

The E-PI concept provides an upper limit on storm velocity in that sense that it invokes a Carnot cycle, which features maximum possible efficiency of converting heat to work. It might therefore appear unsurprising that the majority of storms never reach their E-PI $v_{\rm max}$, with the observed maximum velocities being on average about 40\% lower than the upper limit \citep[e.g.,][Fig.~4b]{sabuwala15}. In contrast,  that some storms, both observed and modeled, significantly exceed their E-PI $V_{\rm max}$ caused more interest and initiated analyses \citep[e.g.,][]{bryan09b,montgomery06}.

Equation~(\ref{JD}) is remarkable as it  relates the {\it local} power of turbulent dissipation in the boundary layer to the {\it local} heat flux from the ocean  via efficiency $\varepsilon_C$, despite the latter is not a local characteristics but applies to a Carnot cycle as a whole. As we discuss below, this local property follows from the dynamic assumptions of E-PI, in particular, from the assumption of gradient wind balance above the boundary layer. \citet{sm08} argued that E-PI
should underestimate maximum velocity as it implicitly applies gradient wind balance to the boundary layer
where this assumption does not hold. \citet{emanuel11} agreed that a correction to $V_{\rm max}$ of about 10-15\% accounting for the boundary layer dynamics is justified, which is approximately what is found in several numerical simulations  \citep[e.g.,][]{wang10,frisius13,kowaleski16}. At the same time, observations of real storms as well as
some models show that the unbalanced flow in the boundary layer can be at least as significant as the balanced flow
and that neglecting these unbalanced effects may cause much more significant underestimates of maximum velocity \citep{bryan09b,montgomery2014,sanger2014}. The question about E-PI underpredicting $V_{\rm max}$ requires a further investigation.

Apart from gradient wind balance discussions, the E-PI concept advanced through several modifications aimed to
remedy the limitations of the original formulation \citep{em88,em91,em95,em97,emanuel11}. A major modification was proposed by \citet{bister98} who suggested that Eq.~(\ref{JD}) should be replaced by $D = \varepsilon_C (J+D)$.
They interpreted this as a situation when turbulent kinetic energy locally dissipates to heat and this heat is added to the
thermodynamic cycle along with the oceanic heat source $J$. This modification of (\ref{JD}) leads to replacement of $\varepsilon_C$ by $\varepsilon_C/(1- \varepsilon_C)$ in (\ref{vmax}), which increases $V_{\rm max}$ by about 20\% for a typical hurricane value of $\varepsilon_C \sim 0.3$ \citep{em86,demaria94}. However, it is not obvious whether turbulent dissipation within the hurricane proceeds down to the thermal level or the ultimate products of this dissipation are small-scale eddies that are exported from the hurricane without contributing to its heat balance. The criteria by which to decide whether thermal dissipation occurs or not are unclear. \citet[][]{kieu15} argued that the dissipative heating can be insignificant. To account for this uncertainty, it became common to estimate $V_{\rm max}$ both with and without dissipative heating \citep[e.g.,][]{montgomery06,bryan09b,sabuwala15}. This created a wide range of predicted values.  E.g. if $V_{\rm max} \sim 70$~m~s$^{-1}$ is a Category 5 hurricane, a 20\% reduction in $V_{\rm max}$ makes it a Category 3.

Recently, \citet{sabuwala15} proposed another significant modification to E-PI suggesting that Eq.~(\ref{JD})
should become $D + W_P = \varepsilon_C (J+D)$, to account for the power $W_P$ needed to lift precipitating water. \citet{sabuwala15} estimated $W_P$ from the observed rainfall in the region of maximum winds. They concluded that accounting for $W_P$ can reduce $V_{\rm max}$ by as much as 30\%. Since theoretical estimates of hurricane $W_P$ appear unavailable \citep[but see][]{pla11}, this proposition created further uncertainty in
E-PI $V_{\rm max}$ values.

Understanding what  determines hurricane intensity is crucial  for improved predictions. Here we re-derive the E-PI concept separating its dynamic and thermodynamic assumptions to address the above three topics --unbalanced flow, dissipative heating and water lifting-- in a comprehensive manner. This requires paying special attention to the assumptions concerning the storm's outflow. In Section~\ref{ePI} we derive E-PI $V_{\rm max}$ (\ref{vmax}) without using gradient wind balance but just assuming the atmosphere above the boundary layer to be inviscid. One of the implications of this result is that, while the radially unbalanced flow is crucial for storm's power, the fact that E-PI underestimates $V_{\rm max}$
may have a different reason. Its discussion is postponed to Section~\ref{cons}. In Section~\ref{mod} we show  that the correct version of Eq.~(\ref{JD}) consistent with E-PI assumptions for the boundary layer  is $D = \varepsilon_C (D+ J_L)$, where $J_L\le J$ is the latent heat flux from the ocean. The presence or absence of dissipative heating does not affect this formulation, but  constrains the sensible heat flux (and vice versa). Furthermore, we show that, contrary to the suggestion of \citet{sabuwala15}, the gravitational power of precipitation $W_P$ makes only a minor impact on hurricane intensity, although this impact grows with decreasing $\varepsilon_C$.

In Section~\ref{cons} we analyze the key dynamic and thermodynamic assumptions of E-PI underlying Eq.~(\ref{vmax}) and show that they are generally not mutually consistent. We demonstrate that this is the reason for a systematic underestimate of maximum velocities by Eq.~(\ref{vmax}) and its published modifications. We discuss how these assumptions can be modified to be compatible.

Finally, in Section~\ref{scale} we perform a scale analysis of key parameters determining $V_{\rm max}$ in E-PI and show that, in agreement with our previous suggestion \citep{ar17}, numerically E-PI implies $\rho V_{\rm max}^2/2 \sim p_{vs}$, where $p_{vs}$ is partial pressure of water vapor in the surface air. We discuss why this result matters.
In Section~\ref{disc} we overview the results and outline ideas requiring further investigation.

\section{Emanuel's Potential Intensity}
\label{ePI}

\subsection{Dynamics}
\label{ePId}

\begin{figure*}[tbp]
\begin{minipage}[p]{0.65\textwidth}
\centering\includegraphics[width=0.8\textwidth,angle=0,clip]{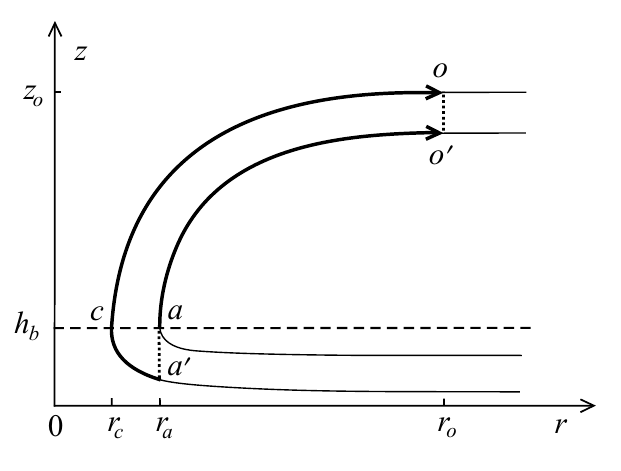}
\end{minipage}
\caption{A schematic illustrating two streamlines leaving the boundary layer $z \le h_b$ in the vicinity of the eyewall.}
\label{fig1}
\end{figure*}

\citet{em86} assumed that the atmosphere is inviscid above the boundary layer $z \le h_b$ and
hydrostatic
\begin{equation}\label{he}
\alpha \frac{\partial p}{\partial z} = -g.
\end{equation}
Consider two streamlines $a' - o$ and $a - o'$ that both begin at $r_{a'} = r_a$  and end at $r_o = r_{o'}$, Fig.~\ref{fig1}.
To these two streamlines we apply the Bernoulli equation
\begin{equation}\label{B}
-\alpha dp = d\left(\frac{V^2}{2}\right) +g dz - \mathbf{f} \cdot d\mathbf{l},
\end{equation}
where $\alpha \equiv 1/\rho$, $p$ is pressure, $V$ is air velocity, $\mathbf{f}$ is the turbulent friction force per unit air mass and $d\mathbf{l}=\mathbf{V}dt$. Additionally using Eq.~(\ref{he})
we find the integral of $-\alpha dp$ over the closed contour $a - c - o - o'-a$ \citep[this contour is considered in~Fig.~8.1 of][]{emanuel_2004}:
\begin{equation}\label{oint}
-\oint \alpha dp = -\int_{a}^c \alpha \frac{\partial p}{\partial r} dr - \frac{V_{c}^2 - V_{a}^2}{2} - \frac{V_{o'}^2 - V_{o}^2}{2}.
\end{equation}
Note that $\mathbf{f} = 0$ for $z > h_b$ and $z_c = z_a = h_b$.
The first term in the right-hand side represents work $A^+$ of the pressure gradient at the top of the boundary layer $z = h_b$. The other two terms represent work $A^-$ above the boundary layer:
\begin{equation}\label{wup}
A^-\equiv -\int\limits_{c-o-o'-a}\alpha dp = -\int\limits_{c-o-o'-a}\alpha \frac{\partial p}{\partial r} dr =
-\left(\frac{V_{c}^2 - V_{a}^2}{2} + \frac{V_{o'}^2 - V_{o}^2}{2}\right).
\end{equation}
This work is equal to the kinetic energy increments along the top of the boundary layer $a-c$ and in the outflow
$o-o'$ taken with the minus sign. As we discuss below, the outflow term is neglected in the E-PI concept.

Equation~(\ref{oint}) is the main dynamic equation of the E-PI concept. Its main
feature is $A^- \le 0$.  Indeed, if $r_c = r_{\rm max} < r_a$, where $r_{\rm max}$ is the radius of maximum wind,
then $V_c > V_a$ and $A^- < 0$. Areas where the generation of kinetic energy by the radial pressure gradient in the upper atmosphere
is negative are indeed noticeable at least in modelled hurricanes -- see, for example, Fig. 42 (panel KB)
of \citet{kurihara75} and Fig. 5a,b of \citet{smith18}.
Equation~(\ref{oint}) indicates that the pressure gradient in the upper atmosphere consumes
all kinetic energy generated in the boundary layer. If points $c$ and $a$ are chosen around the
radius of maximum wind $r_c \le r_{\rm max} \le r_a$  such that $V_c = V_a$, then $A^- = 0$ and
the total work around the contour is simply equal to work $A^+$  at the top of the boundary layer.

\subsection{Thermodynamics}

\citet{em86} assumed that $a-c$ with $T_a = T_c$ and $o-o'$ with $T_o = T_{o'}$ are reversible (i.e. saturated) isotherms, while  $c-o$ and $o'-a$ are reversible adiabates\footnote{We note that \citet{em86} did employ reversible thermodynamics even if he formally ignored the liquid water mass, as pointed out by \citet{bryan09b} and \citet{montgomery17b}. The reversibility consists in the fact that when one follows path $c - o$ up and path $o' - a$ down, which are both pseudoadiabates, one finds that the water vapor first condenses (disappears) along $c - o$ but then evaporates (reappears) along $o' - a$. Without this reappearance of water vapor, path  $o' - a$ were not a moist pseudoadiabat but had near zero humidity. Thermodynamically, this is equivalent to a situation when the liquid water travels along the cycle $a - c - o - o' - a$ together with the air, but has zero mass and zero heat capacity. 
Since the efficiency of the Carnot cycle does not depend on the heat capacity of the substances undergoing the cycle, for the relationship between heat input and total work using two pseudoadiabates in the cycle is the same as using two reversible adiabates.}. In this case  the contour $a- c - o - o' - a$ is a Carnot cycle with efficiency $\varepsilon = \varepsilon_C = (T_a - T_o)/T_a$:
\begin{equation}\label{wc}
-\oint \alpha_d dp = \varepsilon_C \int_{a}^c \delta Q = \varepsilon_C \int_{a}^c T ds^* = \varepsilon_C \int_{a}^c \left(-\alpha_d \frac{\partial p}{\partial r} + L \frac{\partial q}{\partial r}\right)dr.
\end{equation}
Here $\alpha_d \equiv 1/\rho_d$ is the specific volume of dry air, $\rho_d$ is dry air density,  $\delta Q$ is heat received per unit dry air mass, $s^*$ is saturated entropy per unit dry air mass, $\rho_v$ is water vapor density,  $q \equiv \rho_v/\rho_d$ is the water vapor mixing ratio, $L$ (J~kg$^{-1}$) is the latent heat of vaporization. In the last equality of (\ref{wc}) we have used the  first law of thermodynamics to describe the isothermal heat input.  Equation (\ref{wc}) indicates that once an imaginary parcel of moist air containing $m_d$ kg of dry air (plus some moisture) has completed
the cycle, total work (in Joules) it has performed is given by (\ref{wc}) multiplied by $m_d$. Since, unlike the dry air mass $m_d$, the water  vapor mass $m_v$ may change during the cycle, the integral equation (\ref{wc}) cannot be written per unit wet air mass $m = m_d + m_v$.  Note that Eq.~(\ref{wc}) relates state variables independent of whether the air actually moves along the contour.

Using the relationship
\begin{equation}\label{id}
\alpha_d = \alpha (1 + q)
\end{equation}
and combining (\ref{oint}) and (\ref{wc}) we find
\begin{equation}\label{f18}
\varepsilon_C \int_{a}^c \left(-\alpha_d \frac{\partial p}{\partial r} + L\frac{\partial q}{\partial r}\right) dr =
-\int_{a}^c \alpha \frac{\partial p}{\partial r} dr - \frac{V_{c}^2 - V_{a}^2}{2} - \frac{V_{o'}^2 - V_{o}^2}{2}
- \oint \alpha q dp.
\end{equation}
\citet{em86} did not discriminate between
$\alpha$ and $\alpha_d$ (\ref{id}) and thus neglected the last term in (\ref{f18}). Otherwise, as shown
in Appendix A, Eq.~(\ref{f18}) is similar to \citet{em86}'s Eq.~18, which underlies E-PI estimates.

Using (\ref{he}) and (\ref{B}) we can express the last term in Eq.~(\ref{f18}) as follows (see Appendix B for calculation details):
%\begin{linenomath*}
%\begin{multline} \label{est}
\begin{equation} \label{est}
- \oint \alpha q dp =
- \int_a^c q\left( \alpha \frac{\partial p}{\partial r} + \frac{1}{2}\frac{\partial V^2}{\partial r} \right) dr
- \frac{1}{2}\left(q_{o'}V_{o'}^2 - q_oV_o^2\right)   % \\
- \oint \frac{V^2}{2} \frac{\partial q}{\partial r} dr  - \oint gz \frac{\partial q}{\partial z} dz.
\end{equation}
%\end{multline}
%\end{linenomath*}

Since (\ref{f18}) and (\ref{est}) are valid for any points $a$ and $c$ on the isotherm $z=h_b$,  we can, as suggested by Dr. Kerry Emanuel (pers. comm.), consider a narrow contour  with $r_a \to r_c$ and $z_{o'} \to z_o$ (Fig.~\ref{fig1}).  With an account of (\ref{est}) and after some re-arrangement, Eq.~(\ref{f18}) then becomes
\begin{gather}
\label{alg}
\varepsilon_C \left(-\alpha \frac{\partial p}{\partial r} + \frac{L}{1+q}\frac{\partial q}{\partial r}\right) \!=\! - \alpha \frac{\partial p}{\partial r} - \frac{1}{2}\frac{\partial V^2}{\partial r} + K_1 + K_2,
\quad   z = h_b,\\     \label{K}
K_1 \equiv \left(gH_P- \frac{V^2 - \overline{V^2}}{2}\right)\frac{1}{1+q} \frac{\partial q}{\partial r},\quad  K_2 \equiv \frac{1+q_o}{1+q}\frac{1}{2}\frac{\partial V_o^2}{\partial r}, \\ \label{HP}
H_P \equiv -\frac{1}{q_c - q_o}\int\limits_c^o z \frac{\partial q}{\partial z} dz,
\quad  \overline{V^2} \equiv -\frac{1}{q_c - q_o}\int\limits_c^o V^2 \frac{\partial q}{\partial r} dr.
\end{gather}

Equation (\ref{alg}) summarizes the energy budget of an infinitely narrow cycle $a-c-o-o'-a$ and  thus provides a relationship between local variables. The left-hand side of Eq.~(\ref{alg}) represents local isothermal heat input per unit mass of moist air \citep[e.g.,][Eq.~3.8.2]{gill82}  multiplied by the thermodynamic efficiency of the Carnot cycle $\varepsilon_C$.

In the right-hand side $K_1$ represents potential and kinetic energy increments associated with phase transitions.
Term $gH_P$ accounts for the energy expended to lift water (the gravitational power of precipitation); $H_P$ is
the mean height where condensation occurs. The gas (water vapor)  arises at the surface and disappears (i.e. condenses), together with its potential energy, at altitude $H_P$. Note  that the atmosphere must raise newly evaporated water vapor, hence $K_1$ is proportional to the radial  gradient $\partial q/\partial r$ of the water vapor mixing ratio, i.e. to evaporation.

The term $(V^2-\overline{V^2})/2$ accounts for the fact that evaporating water vapor is formally added to the air mixture with velocity $V$ of the air at $a-c$, and then it disappears during condensation with velocity $\overline{V} < V$ in the upper atmosphere. In what is to follow $(V^2 - \overline{V^2})/2$ is neglected in $K_1$ (\ref{K}) as $V^2/2$ is small compared to $g H_P$: since typical mean condensation heights in the tropics are about $H_P \sim 5$~km \citep[][their Fig.~1]{jas13}, with $V \sim 60$~m~s$^{-1}$, $V^2/2$ is only about 4\% of $gH_P$.  (Term $(V^2-\overline{V^2})/2$ can be explicitly accounted for by specifying the interaction between condensate and air (i.e. introducing a specific term to the equations of motion and Bernoulli equation).  If condensate is assumed to have the same horizontal velocity as air
\citep[see, e.g.,][]{oo01,jgra17}, then as it leaves the air at the surface it has the same velocity as the newly evaporated water vapor and thus net impact of this term to the power budget will be zero. An explicit account of this effect is somewhat lengthy \citep[but see][Fig.~1]{arxiv17}.)

The term $K_2$ describes how the kinetic energy in the outflow, $V_o^2/2$, changes depending on the radius $r=r_c$ where the streamline crosses the top of  the boundary layer $z = h_b$, Fig.~\ref{fig1}.  Point $(r_o,z_o)$, to which $q_o$ and $V_o$ refer, is defined by the condition $\partial T/\partial z = 0$.
(Note that by chain rule $(\partial V_o/\partial z) dz |_{r = r_o} =  -(\partial V_o/\partial r) dr|_{z = h_b}$, see also Appendix C.)

\subsection{Transition from work to power}

If $a-c$ were a streamline, the Bernoulli equation (\ref{B}) could be applied and   the first two terms in the right-hand side of (\ref{alg}) times $dr$ would be equal to turbulent friction $-\mathbf{f}\cdot d\mathbf{l}$.  In reality, however, the air converges towards the windwall at $z < h_b$ and ascends in the boundary layer, such  that every streamline leaves $z = h_b$ at its own $r$, Fig.~\ref{fig1}.  Using (\ref{he}) and applying the Bernoulli equation to part $a' - c$ of the $a'-o$ streamline we find in the limit $r_{a'} \to r_c$, $z_{a'} \to z_c = h_b$:
\begin{equation}\label{fdl}
-\mathbf{f}\cdot d\mathbf{l} = -\alpha \frac{\partial p}{\partial r} dr- \frac{1}{2}\frac {\partial V^2}{\partial r} dr - \frac{1}{2}\frac {\partial V^2}{\partial z} dz,\quad z=h_b.
\end{equation}
We can set the last term in the right-hand side of Eq.~(\ref{fdl}) to zero by assuming, as done in E-PI,
that at any $r$ the velocity reaches its maximum at $z=h_b$ \citep[][p.~3045]{bryan09b}. Then, multiplying Eqs.~(\ref{fdl}) and ~(\ref{alg}) by $\rho$ and $\rho dr$, respectively, and dividing both equations by $dt$ we obtain
\begin{equation}\label{pow}
\varepsilon_C \left(-\frac{\partial p}{\partial r} + \frac{L\rho}{1+q} \frac{\partial q}{\partial r}\right) u
= -\rho \mathbf{f}\cdot \mathbf{V} + (K_1 + K_2) \rho u, \quad  z= h_b,
\end{equation}
Here $u \equiv dr/dt$ is radial velocity and $\mathbf{V} \equiv d\mathbf{l}/dt$ is total air velocity;
$-\rho \mathbf{f}\cdot \mathbf{V}$ (W~m$^{-3}$) is the volume-specific rate of turbulent dissipation
in the air parcels leaving the boundary layer.

If we additionally assume, as done in E-PI \citep[see][Fig.~1]{em86}, that the reversible adiabates in the boundary layer are vertical,  particularly that $\partial s^*/\partial z = 0$ at $z = h_b$, then the term in parentheses in the left-hand side of Eq.~(\ref{pow}) acquires the meaning of the volume-specific rate of heat input into the air parcels leaving the boundary layer. Indeed, in this case heat input into the air parcels moving along $a'-c$  only depends on the horizontal gradients of state variables; this heat input is isothermal in the considered limit $z_{a'} \to h_b$, since the top of the boundary layer is assumed to be isothermal.

\subsection{Boundary layer}

We recapitulate that Equation~(\ref{pow}) relates the volume-specific rates of turbulent dissipation and heat input  at $z= h_b$ and $r = r_c$ under the following assumptions: $c - o$ is an inviscid reversible adiabat  in a hydrostatic atmosphere; immediately below $z = z_c = h_b$ the reversible adiabat is vertical;  $\partial V/\partial z = 0$ and $\partial T/\partial r = 0$ at $z = h_b$, $\partial T/\partial z = 0$ at $r = r_o$.  Notably, the value of $r_o$ should be defined by the condition $\partial T/\partial z = 0$. E-PI presumes  that $r_o$ additionally conforms to the condition that $K_2$ in (\ref{pow}) is negligible. (There is no {\it a priori}  guarantee that the two conditions are compatible, see Section~\ref{cons}). For example,  \citet{bister98} consider the case when $r_o$ is chosen such that $V_o = v_o = 0$, where  $v_o$ is tangential velocity at point $o$.  The gravitational power of precipitation, which derives from the last term in Eq.~(\ref{f18}), is also neglected, $K_1 = 0$.

With $K_1 = K_2 = 0$, the E-PI version of Eq.~(\ref{pow}) tells us that at the top of the boundary layer
the volume-specific rate of turbulent dissipation  is $\varepsilon_C$ times the volume-specific rate of heat input into the local air parcels.  We now assume that these two rates $d$ and $j$ (W~m$^{-3}$) relate as the corresponding surface fluxes (W~m$^{-2})$ of turbulent  dissipation $D$ (\ref{D}) and oceanic heat input $J$ (\ref{J}):
\begin{equation}\label{dj}
\frac{d}{j} = \frac{D}{J},
\end{equation}
where
\begin{equation}\label{djdef}
d = -\rho \mathbf{f}\cdot \mathbf{V}, \quad  j = \left(-\frac{\partial p}{\partial r} + \frac{L\rho}{1+q} \frac{\partial q}{\partial r}\right) u.
\end{equation}
\noindent
Assumption (\ref{dj}) has the following meaning. In E-PI
the atmosphere above the boundary layer is assumed to be inviscid (no turbulent dissipation)
and the air motion adiabatic (no heat input).
All surface flux of turbulent dissipation $D$ (\ref{D}) and all surface heat flux $J$ (\ref{J})
are accommodated into the air within the boundary layer $z \le h_b$.
The mean volume-specific rates of oceanic heat input $\overline{j}$
and turbulent dissipation $\overline{d}$ in the boundary layer at a given $r$ are
$\overline{j}(r) = J/h_b$ and $\overline{d}(r) = D/h_b$.

Then, if the volume-specific rates of oceanic heat input $j$ and turbulent dissipation $d$
at the top of the boundary layer represent equal fractions $\kappa$ of the corresponding mean values,
\begin{equation}\label{djm}
j = \kappa \overline{j}(r) = \kappa\frac{J}{h_b},\quad d = \kappa\overline{d}(r) = \kappa \frac{D}{h_b},
\end{equation}
Eq.~(\ref{dj}) follows. In the same context,  \citet{emanuel11} emphasized that in E-PI  the turbulent fluxes of heat and momentum should have one and the same scale height. Strictly speaking, Eq.~(\ref{djm}) additionally assumes that the horizontal turbulent heat flux is negligible (i.e. all heat added to the air at a given $r$ is of local origin).  Furthermore, defining the volume-specific rate of turbulent dissipation as $d \equiv -\rho\mathbf{f}\cdot \mathbf{V}$ in Eq.~(\ref{djdef}) assumes that the divergent part of the rate of work of turbulent friction \citep[the last but one term in Eq.~7 of][]{fiedler00} is negligible, i.e. $-\rho\mathbf{f}\cdot \mathbf{V}$
is positive definite as is $D$ (\ref{D}).

Using Eqs.~(\ref{D}), (\ref{J}), (\ref{dj}) and (\ref{pow})  (with $K_1 = K_2 = 0$ in the latter)  we find
\begin{equation}\label{V}
V^2 = \frac{D}{J} \frac{C_k}{C_D} (k_s^* - k) = \frac{d}{j} \frac{C_k}{C_D} (k_s^* - k) = \varepsilon_C \frac{C_k}{C_D} (k_s^* - k).
\end{equation}
It is noteworthy that deriving Eq.~(\ref{V}) does not require  $\partial V/\partial r = 0$ (the condition of maximum wind) \citep[this peculiarity of E-PI was noted by][]{montgomery17b}.

\subsection{Balanced versus unbalanced flow}

Unlike \citet{em86}, in our derivation of Eq.~(\ref{V}) we have not assumed that above the boundary layer
the atmosphere is in gradient wind balance (see Appendix A).  Our derivation is more general: we have only assumed that the atmosphere is inviscid.  Velocity $V$ in (\ref{V}) is total air velocity.

Likewise our assumption $d/j = D/J$ (\ref{dj}) is more general than E-PI's
\begin{equation}\label{Ms}
\frac{\partial s^*}{\partial M} \bigg|_{z=h_b} = \frac{\tau_s}{\tau_M} \bigg|_{z=0},
\end{equation}
 where
$s$ and $M$ and $\tau_s$, $\tau_M$ are entropy and angular momentum and their vertical surface fluxes, respectively \citep[see][Eq.~32]{em86}.  Assumption (\ref{Ms}) is meaningful when the top of the boundary layer is in gradient wind balance.  Indeed,  assuming that in gradient wind balance $V^2 \approx v^2$, where $v$ is tangential velocity,  and that in the region of maximum winds (small radii) $v \approx M/r$, where $M$ is angular momentum,  we find from Eq.~(\ref{fdl}) (as before with $\partial V/\partial z = 0$):
\begin{equation}\label{fdl2}
-\mathbf{f}\cdot d\mathbf{l}
\approx  - \frac{M}{r^2}\frac{\partial M}{\partial r}dr + \left(\frac{v^2}{r} -\alpha \frac{\partial p}{\partial r} \right) dr,
\quad z = h_b,  \quad  r = r_{\rm max}.
\end{equation}
In gradient wind balance the expression in parentheses in Eq.~(\ref{fdl2}) is zero. Only in this case turbulent dissipation $-\mathbf{f}\cdot d\mathbf{l}$ describes the radial change of angular momentum, while Eq.~(\ref{pow}) (with $K_1 = K_2 = 0$) relates  heat input $T_a \partial s^*/\partial r$ to $\partial M/\partial r$ and becomes \citet{em86}'s Eq.~13:
\begin{equation}\label{s}
(T_a - T_o) \frac{\partial s^*}{\partial r} = -\frac{M}{r^2} \frac{\partial M}{\partial r}.
\end{equation}
In gradient wind balance assumptions (\ref{dj}) and (\ref{Ms}) are equivalent, since $\tau_s = J/(\rho T_s)$ and $\tau_M = -C_D r V^2=-r^2D/(\rho M)$, cf. Eqs.~(\ref{D}), (\ref{J})  and \citet{em86}'s Eq.~33. In this case, with $T_s \approx T_a$, Eqs.~(\ref{Ms}) and (\ref{s}) yield Eq.~(\ref{V}) as  do our more general Eqs.~(\ref{dj}) and (\ref{pow}).

However, when the flow is strongly unbalanced and the term in parentheses in (\ref{fdl2}) is significant, there is no
proportionality between the change of momentum and turbulent dissipation in  Eq.~(\ref{fdl2}). Assumption~(\ref{Ms}) is no longer logical. For example, in the limit when the flow is strictly radial,  the angular momentum is about zero
and $\partial M/\partial r \approx 0$ at any sufficiently small $r$. But turbulent dissipation and heat input $T_a \partial  s^*/\partial r$ in the radial flow are not zero  and Eq.~(\ref{pow}) remains valid.  Thus, while neither Eq.~(\ref{dj}) nor Eq.~(\ref{Ms}) are particularly rigorously justified, in the strictly radial flow Eq.~(\ref{dj}) can  be valid, while Eq.~(\ref{Ms}) definitely cannot\footnote{\citet{bryan09b} use 
Eq.~(\ref{Ms}), rather than our Eq.~(\ref{dj}), for the radially unbalanced flow, hence the difference between
their Eq.~24 and our formulations. Another difference consists in the fact that we consider a radially unbalanced
hydrostatic flow, while \citet{bryan09b} consider a flow that is neither radially nor hydrostatically balanced.
However, \citet{bryan09b} point out that in the location of maximum tangential wind, to which their Eq.~24 for revised potential intensity refers, the vertical velocity $w$ is also maximum. In this case, the hydrostatic
equilibrium locally holds.}.

In both models and observations E-PI $v_{\rm max}$ (\ref{vmax}) sometimes underestimates the actual maximum velocity \citep[e.g.,][]{bryan09b,montgomery06}.  At the same time, several empirical and modelling studies indicate that gradient wind balance can be significantly perturbed both  in the boundary layer and above it  \citep{sm08,bryan09b,montgomery2014,sanger2014}. The two patterns were suggested to be related: E-PI Eq.~(\ref{vmax}) underestimates the actual maximum velocity in the unbalanced flow because it derives from gradient wind balance. This appears plausible: in the flow converging towards the eyewall the centrifugal force $v^2/r$ is smaller than the inward pressure gradient force,  thus the term in parentheses in Eq.~(\ref{fdl2}) is positive. Neglecting this term under the assumption of gradient wind balance underestimates turbulent dissipation $-\mathbf{f}\cdot d\mathbf{l}$ and $D$ and thus ultimately $v_{\rm max}$ calculated from Eq.~(\ref{vmax}).

However, we have shown that in the radially unbalanced flow the expression for maximum velocity (\ref{V}) is identical
to E-PI Eq.~(\ref{vmax}) derived assuming gradient wind balance and $V = v$.  This suggests that the mismatch between Eq.~(\ref{V})  and observations may have a different explanation. Indeed,  the mismatch between $V_{\rm max1}$ and $V_{\rm max2}$ values derived from Eq.~(\ref{fdl2}) preserving (1) or discarding (2) the term in parentheses is not the same as the mismatch between $V_{\rm max1}$ and observations. We consider the latter discrepancy
in Section~\ref{cons}.

\section{Modifications to E-PI}
\label{mod}

\subsection{Sensible heat, latent heat and dissipative heating}
\label{dh}

\citet{bister98} proposed that besides the ocean there is another source of heat -- the kinetic energy of wind which locally dissipates to heat.  In this case the volume-specific rate of heat input as described  by the first law of thermodynamics in the left-hand side of Eq.~(\ref{pow}) can be written as the sum of two sources, heat $j$ from the ocean and heat $j_D \le d$ from turbulent dissipation. In the presence of such dissipative heating Eq.~(\ref{djdef}) becomes
\begin{equation}\label{djdef1}
d = -\rho \mathbf{f}\cdot \mathbf{V},\quad j + j_D = \left(-\frac{\partial p}{\partial r} + \frac{L\rho}{1+q} \frac{\partial q}{\partial r}\right) u.
\end{equation}
If all turbulent dissipation proceeds down to the thermal level locally within the storm, then $d = j_D$,
and from Eq.~(\ref{pow}) (with $K_1 = K_2 = 0$) and Eqs.~(\ref{dj}) and (\ref{djdef1}) we obtain
\begin{equation}\label{be}
D = \frac{\varepsilon_C}{1 - \varepsilon_C} J,\quad
V^2 = \frac{\varepsilon_C}{1 - \varepsilon_C} \frac{C_k}{C_D} (k_s^* - k) = \frac{\varepsilon_C}{1 - \varepsilon_C} \frac{C_k}{C_D}(c_p \Delta T + L \Delta q),
\end{equation}
which is the main result of \citet[][see their Eq.~21]{bister98}.

While it appears that dissipative heating elevates hurricane intensity due to the factor $1/(1-\varepsilon_C)> 1$,
scrutinizing Eqs.~(\ref{pow}) and (\ref{djdef1}) reveals that dissipative heating cannot be added without reducing sensible heat flux by a similar amount as previously noted by \citet[][]{dhe10} \citep[see also][]{bejan19}. Using, respectively, Eqs.~(\ref{J}) and (\ref{djdef1}) the surface-specific and the volume-specific oceanic heat fluxes $J$ and $j$ can be represented as a sum of the latent heat flux $J_L$ and $j_L$ (associated with the mass flux of water vapor) and sensible heat flux $J_S$ and
$j_S$:
\begin{gather}\label{JLS}
J = J_L + J_S,  \quad J_L = \rho C_k V L \Delta q, \quad  J_S =  \rho C_k V c_p \Delta T, \\
j = j_L + j_S,  \quad j_L = \frac{L\rho }{1+q}\frac{\partial q}{\partial r}u, \quad  j_S = -\frac{\partial p}{\partial r}u - j_D.  \label{jls}
\end{gather}
\noindent
Here $j_L$ describes the turbulent admixture into air parcels of water vapor evaporated from the ocean,
while $-u\partial p/\partial r$ describes the rate at which an expanding air parcel that performs work must
receive heat to remain isothermal. This heat can derive
from the oceanic flux of sensible heat $j_S$ governed by the temperature difference between the ocean
and the adjacent air as well as from the local thermal dissipation rate $j_D$ of kinetic energy.

Considering (for the first time in our derivations) that at the radius of maximum wind
$\partial V/\partial r= 0$, Eq.~(\ref{fdl}) divided by $dt$ and multiplied by $\rho$ gives
\begin{equation}\label{turb}
d = -\rho \mathbf{f}\cdot \mathbf{V} = -\frac{\partial p}{\partial r}u .
\end{equation}

If $d= j_D$ (complete thermal dissipation of turbulence), then from Eqs.~(\ref{turb}) and (\ref{jls})
we find $j_S = 0$, i.e. the sensible heat flux is zero. Total heat flux $j$ keeps the air isothermal by compensating for the loss of internal energy that would otherwise occur as the air performs work. At the radius of maximum wind the rate of this work equals the rate of its turbulent dissipation, as shown by Eq.~(\ref{turb}). The dissipative heating from turbulence
can exactly compensate for the loss of internal energy to work -- and thus sensible heat is no longer required nor can it be accomodated by the air parcel without its temperature rising. In other words, Eq.~(\ref{turb}) constrains the sum of sensible and dissipative heat fluxes, $j_S + j_D = d$.

\citet{kieu15} reached similar conclusions and showed that the dissipative heating is inherently included in the power budget of the hurricane and cannot be treated as a separate heat source. (Unlike in the present work, \citet[][see his Eq.~13]{kieu15}  did not analyze the local equations of E-PI but used integral equations for the hurricane as a whole.)
In numerical models where dissipative heating is included as an additional term in the power budget
the air becomes warmer than the ocean producing such exotic processes as a large negative flux of sensible heat at peak hurricane intensity \citep{zhang1999}. (This contradicts observations.) The increase in maximum velocity in such models arises from a larger $\Delta q$ rather than from the factor $1/(1-\varepsilon_C)>1$ in Eq.~(\ref{be}).

Using Eq.~(\ref{turb}) in Eq.~(\ref{pow}) with $K_1 = K_2 = 0$ and taking into account Eq.~(\ref{jls})
we find
\begin{equation}\label{djl}
d = \frac{\varepsilon_C}{1 - \varepsilon_C} j_L.
\end{equation}
This means that in E-PI the flux of turbulent dissipation at the radius of maximum wind is unambiguously
related to the latent heat flux (and not to the total oceanic heat flux $j$). Assuming by analogy
with Eq.~(\ref{djm}) that $j_L = \kappa J_L/h_b$, $j_S = \kappa J_S/h_b$ and thus
\begin{equation}\label{dj2}
\frac{j_S}{d} = \frac{J_S}{D},\quad \frac{j_L}{d} = \frac{J_L}{D},
\end{equation}
we obtain, instead of Eq.~(\ref{be}),
\begin{equation}\label{beL}
D = \frac{\varepsilon_C}{1 - \varepsilon_C} J_L,\quad
V_{\rm max}^2 = \frac{\varepsilon_C}{1 - \varepsilon_C} \frac{C_k}{C_D}L \Delta q.
\end{equation}

Essentially, unlike Eq.~(\ref{V}), which assumes $j_D = 0$ (i.e., no dissipative heating, $j_S = d$)
or Eq.~(\ref{be}), which assumes $j_D = d$ (complete thermal dissipation of turbulence, $j_S = 0$),
Eq.~(\ref{beL}) is valid independent of how $j_D$ and $d$ relate.  Irrespective of whether turbulent dissipation contributes something  to heat balance (or all turbulent energy is exported from the storm in the form
of small eddies), Eq.~(\ref{beL}) relates turbulent dissipation to oceanic {\it latent} heat flux at the radius of maximum wind. This constitutes a remarkable and convenient feature of E-PI: it is independent of the parameter $\Delta T$ (\ref{JLS})  governing the sensible heat flux from the ocean. Accordingly, neither $\Delta T$, nor sensible versus latent
heat fluxes have previously been assessed theoretically within the E-PI framework.

In the general case, when the fraction $\eta$ of turbulent flux $d$ undergoes thermal dissipation locally
(and the remaining part of small-scale eddies are exported from the storm), i.e. $j_D = \eta d$,
we find from Eq.~(\ref{pow}) (with $K_1 = K_2 = 0$), Eqs.~(\ref{djdef1}), (\ref{dj2}) and  (\ref{turb})
\begin{equation}\label{JSn}
J_S = (1 - \eta) \frac{\varepsilon_C}{1 - \varepsilon_C} J_L,\quad J = J_S + J_L = \frac{1 -\eta \varepsilon_C}{1 - \varepsilon_C} J_L.
\end{equation}
Thus, under E-PI assumptions, the observed Bowen ratio $J_S/J_L$ can inform us
about the degree $\eta$ to which thermal dissipation occurs at the radius of maximum wind.
For example, a typical ratio $J_S/J_L$ of about one third and a typical $\varepsilon_C = 1/3$
imply $\eta = 1/3$.  Equation~(\ref{JSn}) shows that the conventional relationships $D = \varepsilon_C J$ (\ref{JD}) and Eq.~(\ref{vmax})
 will underestimate $V_{\rm max}$ (\ref{beL}) by $\eta \varepsilon_C$.

In contrast, since Eq.~(\ref{be}) implies $\eta = 1$ and thus $J_S = 0$, applying
Eq.~(\ref{be}) to storms where $J_S > 0$ will overestimate $V_{\rm max}$ (\ref{beL}). This has implications
for studies exploring why E-PI may underpredict $V_{\rm max}$.
In particular, \citet[][see their Eq.~10]{bryan09b} used Eq.~(\ref{be}) as the theoretical E-PI limit.
This is equivalent to comparing their model's maximum intensity $V^2_m$ to $V_{\rm max}^2(J_L+J_S)/J_L$.
Thus, while \citet[][see their Fig.~12]{bryan09b} concluded that $V_{\rm max}^2(J_L+J_S)/J_L = V^2_m$,
the theoretically predicted $V_{\rm max}^2$ (\ref{beL}) with $J_S > 0$ still underestimates $V^2_m$ even
though the unbalanced effects have been taken into account. In Section~\ref{cons} we discuss
possible reasons.

\subsection{Correction due to lifting water}
\label{sHP}
We will now estimate the significance of the so far neglected term $K_1$ in Eqs.~(\ref{alg}) and (\ref{pow}).
Combining $K_1 \rho u$ with $j_L$ (\ref{jls}) in Eq.~(\ref{pow}) (with $K_2 = 0$) and
using Eqs.~(\ref{turb}) and (\ref{dj2}) we obtain
\begin{gather}\label{D0}
\varepsilon_C (D + J_L) = D + g H_P E,\\
\label{D1}
D = \frac{\varepsilon_C - g H_P/L}{1-\varepsilon_C} J_L.
\end{gather}
\noindent
Here $E \equiv J_L/L$ (kg~m$^{-2}$~s$^{-1}$) is the local flux of evaporation.

Our revised E-PI estimate of maximum velocity follows from (\ref{D1}), (\ref{JLS}) and (\ref{D}):
\begin{equation}\label{vmax1}
V_{\rm max}^2 = \frac{\varepsilon_C - g H_P/L}{1-\varepsilon_C} \frac{C_k}{C_D} L\Delta q.
\end{equation}

Mean condensation height $H_P$ (\ref{HP}) can be calculated from the equation of moist adiabat and depends on surface temperature,  the incompleteness of condensation (i.e. altitude where condensation ceases) and, to a lesser degree, on surface relative humidity.  In the tropics $H_P$ does not exceed $6$~km \citep[see][their Fig.~1]{jas13}.
With $H_P = 6$~km we have $gH_P/L = 0.024$ in (\ref{vmax1}).  For a typical $\varepsilon_C \approx 0.3$ the gravitational power of precipitation reduces  $V_{\rm max}$ by approximately $(gH_P/L)/\varepsilon_C/2 \approx 0.04$, i.e. by 4\% at maximum, although this  correction would grow with decreasing $\varepsilon_C$.
(Dr. Kerry Emanuel (pers. comm.) suggested that the ultimate correction would be even smaller if one takes into account the heat associated with dissipation of falling hydrometeors.) This contrasts with the results of \citet{sabuwala15} who found that such a reduction can reach as much as 30\%, with an average of 20\%.

\citet[][their Eq.~(2), the ``adiabatic case'']{sabuwala15} based their analysis
 on the equation
\begin{gather}\label{DS0}
\varepsilon (D + J) = D + g H_P P,\\ \label{DS1}
D = \frac{\varepsilon_C - g H_P/L (PL/J) }{1-\varepsilon_C} J.
\end{gather}
\noindent
Here $P$ (kg~m$^{-2}$~s$^{-1}$) is the local precipitation in the region of maximum winds. Equation~(\ref{DS1}) is obtained  from Eq.~(2) of \citet{sabuwala15} and their additional equation $\dot{Q}_{in} - \dot{Q}_{out} = P$. Note the following differences in notations between \citet{sabuwala15} and the present work:  $T_s \to T_a$, $\dot{Q}_{in} \to J$,
$\dot{Q}_{d}\to D$, $P \to W_P = g H_P P$ \citep[for the last relationship see][]{pa00,jas13}.

Comparing Eqs.~(\ref{DS0}) to (\ref{D0}) we can see that \citet{sabuwala15} used local precipitation $P$ instead of local evaporation $E$ and total heat flux $J$ instead of latent flux $J_L$. The latter difference did not considerably affect the relative magnitude of the correction, but the former one did. Replacing $E$ by $P$ in (\ref{DS0}) as compared to (\ref{D0}) leads to the appearance of a large factor $PL/J \gg 1$ at the gravitational term $g H_P/L$ in (\ref{DS1}) as compared to (\ref{D1}). For typical Bowen ratios in hurricanes $B \equiv J_S/J_L \approx 1/3$ \citep[e.g.,][]{jaimes15}
we have $J = (1 + B)J_L = (1+B) EL$ and $PL/J = (P/E)/(1+B)$.

Ratio $P/E$ between local precipitaton and evaporation in the region of maximum winds is variable but on
average of the order of $10$ \citep[see][their Table~1 and Figs.~2 and 3]{ar17}. The corresponding
correction to $V_{\rm max}$ of \citet{sabuwala15} is $10 (gH_P/L)/[2(1+B) \varepsilon_C] \approx 0.3$, which is almost an order of magnitude larger  than in our Eq.~(\ref{vmax1}), i.e. about 30\%.

\citet{sabuwala15} did not derive their formulations from the orignial assumptions of E-PI.  In their formulation they included what appeared to be a plausible term describing the gravitational power of precipitation
to the hurricane's power budget. However, as our analysis has shown,  the unjustified replacement of evaporation $E$ by precipitation $P$ caused the estimate of \citet{sabuwala15} be too high.

Why does the local power budget of a hurricane in E-PI include evaporation rather than rainfall? Much of the water precipitating within a hurricane is imported from outside \citep{ar17}. The E-PI does not explicitly account for this imported moisture. It views the hurricane as a steady-state thermodynamic cycle with all moisture provided locally by evaporation from the ocean. Imported moisture is, however, implicitly accounted for by considering the hypothetical adiabat $o'-a$ to be part of the hurricane's thermodynamic cycle (Fig.~\ref{fig1}).

Along $o'-a$ the moisture content of the hypothetically descending air rises by over two orders of magnitude.
Most moisture condensing within the hurricane precipitates and cannot serve as a source of water vapor
for the descending air. The vertical distribution of humidity along the $o'-a$ path is provided by evaporation and convection outside the storm. This moisture is gathered by the hurricane as it moves through the atmosphere. As the moisture arrives with its own gravitational energy, the storm spends no energy to raise this imported water. The storm only has to expend energy to raise locally evaporated water, which is why it is evaporation $E$ that enters the power budget equation (\ref{D0}). Notably, \citet{em88} indicated that the term accounting for water lifting energy is proportional
to the difference in the water profiles of the air rising along $c-o$ path and the environmental air along the hypothetical path $o'-a$ \citep[][see his Appendix C and Eq.~C12]{em88}. This difference is indeed provided by evaporation along the path $a-c$.

\section{Consistency of E-PI assumptions}
\label{cons}

\subsection{Isothermy and saturation at $z = h_b$}

We have discussed that in E-PI the top of the boundary layer (path $a-c$) is assumed to be saturated and isothermal. Only in this case the cycle can be reversible and feature Carnot efficiency, such that Eq.~(\ref{wc}) is valid. We will now show that this assumption is mathematically incompatible with the assumption $K_2 = 0$ in Eq.~(\ref{alg}), which underlies the expression for E-PI intensity (\ref{vmax}).

From the definition of $q$ and the ideal gas law
\begin{equation}\label{qdef}
q \equiv \frac{\rho_v}{\rho_d} = \frac{M_v}{M_d} \frac{p_v}{p_d} = (1+q)\frac{M_v}{M} \frac{p_v}{p} = (1+q)
\frac{M_v}{M} \frac{p_v^*}{p}\mathcal{H},
\end{equation}
where $M_v$, $M_d$ and $M$ are the molar masses of water vapor, dry air and air as a whole, $p_v^*$ is the saturated partial pressure of water vapor and $\mathcal{H}$ is relative humidity we have
\citep[for derivation details see][Eq.~3]{ar17}:
\begin{equation} \label{qs}
\frac{dq}{q} = \frac{1}{1-\gamma} \left(- \frac{dp}{p} +\frac{dp^*_v}{p^*_v} + \frac{d\mathcal{H}}{\mathcal{H}} \right)
= (1+q)\frac{M_d}{M} \left(- \frac{dp}{p} +\frac{\mathcal{L}}{RT}\frac{dT}{T} + \frac{d\mathcal{H}}{\mathcal{H}} \right),
\end{equation}
where $\gamma \equiv p_v/p$, $\mathcal{L} = L M_v = 45$~kJ~mol$^{-1}$ is molar heat of vaporization,
$R = 8.3$~J~K$^{-1}$~mol$^{-1}$ is the universal gas constant. In the second equality we have used Clausius-Clapeyron law. We emphasize that Eq.~(\ref{qs}) derives from the definition of $q$, the ideal gas law and Clausius-Clapeyron law and does not involve any assumptions.

Using the ideal gas law in the form $p = (\rho/M) RT$ we can write Eq.~(\ref{qs}) for a horizontal path as
\begin{equation}\label{two}
(1-\gamma)\frac{RT}{\mathcal{L} \gamma} \frac{L}{1+q} \frac{\partial q}{\partial r}= -\alpha \frac{\partial p}{\partial r} +\frac{RT}{M}\left(\frac{\mathcal{L}}{RT^2} \frac{\partial T}{\partial r} + \frac{1}{\mathcal{H}} \frac{\partial \mathcal{H}}{\partial r}\right).
\end{equation}
On the other hand, from Eq.~(\ref{alg}) at the radius of maximum wind, where $\partial V/\partial r = 0$, we obtain, cf. Eq.~(\ref{djl}):
\begin{equation}\label{one}
\frac{\varepsilon_C}{1 - \varepsilon_C} \frac{L}{1+q}\frac{\partial q}{\partial r} = -\alpha \frac{\partial p}{\partial r} +
\frac{K_1+K_2}{1-\varepsilon_C}.
\end{equation}

Comparing Eqs.~(\ref{one}) and (\ref{two}) we find that along an isothermal $\partial T/\partial r = 0$ and saturated $\partial \mathcal{H}/\partial r = 0$ path,
like $a-c$ is assumed to be in E-PI, with $K_1 = K_2 = 0$ these equations cannot be valid simultaneousy unless
\begin{equation}\label{C}
\mathcal{C}_1 \equiv \frac{\varepsilon_C}{1 - \varepsilon_C} = (1-\gamma)\frac{RT}{\mathcal{L}\gamma} \equiv \mathcal{C}_2.
\end{equation}
These conditions are, however, never met in the terrestrial atmosphere. The {\it maximum} value $\mathcal{C}_1 = 0.54$
corresponds to the maximum observed Carnot efficiency $\varepsilon = 0.35$ \citep{demaria94}. The {\it minimum} value $\mathcal{C}_2= 1.1$ is twice as large; it corresponds to the largest $\gamma = p_v^*/p \lesssim 0.05$ (for $T_a = 303$~K and $p = 900$~hPa).

This result means that {\it irrespective of whether the atmosphere above the boundary layer conforms to E-PI assumptions}, if the atmosphere does conform to the boundary layer assumptions (\ref{dj}), (\ref{djdef}) and (\ref{djl}), (\ref{jls}), (\ref{JLS}) then our revised E-PI estimate~(\ref{vmax1}) will systematically {\it underestimate} the observed $V_{\rm max}^2$ by at least a factor of two, $\mathcal{C}_2/\mathcal{C}_1\ge 2$. The conventional estimate (\ref{be}), which uses total heat flux $J$ instead of latent heat flux $J_L$, while unjustified, will display a smaller but still significant disagreement with observations: for $J \approx 1.3 J_L$ we have $(J_L/J)(\mathcal{C}_2/\mathcal{C}_1) \ge 1.5$. This implies a 50\% and 25\% underestimate of observed $V_{\rm max}$ by, respectively, Eqs.~(\ref{beL}) and (\ref{be}).
Similar figures have been reported in the literature \citep[][and references therein]{persing2003,hausman06,cram2007,yang2007,bryan09b}.

\subsection{Implications of $\mathcal{C}_1 \ne \mathcal{C}_2$}
\label{impl}

The discrepancy between Eqs.~(\ref{one}) and (\ref{two}) indicates that the four E-PI assumptions, $\partial T/\partial r = 0$, $\partial \mathcal{H} /\partial r = 0$ at $z = h_b$ and $\partial T/\partial z = 0$, $K_2 = 0$ at $z = z_o$ are mutually inconsistent. (We have shown in Section~\ref{sHP} that $K_1$, the gravitational power of precipitation, is relatively small
and cannot resolve the significant discrepancy, so for simplicity as before we assume $K_1 = 0$.) Even if a radius $r_o$ exists where $\partial T/\partial z = 0$ (such that $o-o'$ is the second isotherm of the Carnot cycle), the kinetic energy change $K_2$ (\ref{K}) at this radius cannot be negligible.

In gradient wind balance $K_2$ is given by Eq.~(\ref{K2}) (Appendix C) and Eq.~(\ref{one}) with $K_1 = 0$
at the radius of maximum wind becomes
\begin{equation}\label{one2}
\frac{\varepsilon_C}{1 - \varepsilon_C} \frac{L}{1+q}\frac{\partial q}{\partial r} =
-\alpha \frac{\partial p}{\partial r} \left(1- \frac{1}{1-\varepsilon_C}\frac{r^2}{r_o^2}\right),
\quad r = r_{\rm max},\quad v = v_{\rm max}.
\end{equation}
(In Eq.~(\ref{one2}) we have neglected a small term $fr/[2v(1-\varepsilon_C)] \ll 1$ as compared to unity, cf. Eq.~(\ref{K2}). Taking conservatively $f = 10^{-4}$~s$^{-1}$, $r_{\rm max} = 40$~km, $v_{\rm max} = 40$~m~s$^{-1}$ and $\varepsilon_C = 0.3$ we have $fr/[2v(1-\varepsilon_C)] <0.1$).

For Eq.~(\ref{one2}) to coincide with Eq.~(\ref{two}), the expression in parentheses in the right-hand
part of Eq.~(\ref{one2}) should be equal to $\mathcal{C}_1/\mathcal{C}_2$. This condition relates the radius of
maximum wind to $r_o$ as
\begin{equation}\label{rmro}
\left(\frac{r_o}{r_{\rm max}}\right)^2 = \frac{1}{(1-\mathcal{C}_1/\mathcal{C}_2)(1-\varepsilon_C)}<\frac{2}{1-\varepsilon_C}.
\end{equation}
If condition $\partial T/\partial z = 0$ is fulfilled on streamline $c-o$ at a radius $r_o$ not significantly exceeding
the radius of maximum wind, Eqs.~(\ref{two}) and (\ref{one2}) can coincide, while Eq.~(\ref{one}) with $K_2 = 0$ will be invalid. Note that in E-PI $r_o$ is defined as a radius where $K_2 = 0$ \citep[e.g.,][p.~588]{em86} (this radius is additionally assumed to feature the property $\partial T/\partial z = 0$). Thus E-PI constraints on $r_o$ and its relationship to $r_{\rm max}$ \citep[e.g., that $r_o/r_{\rm max} > 10$, see][his Table~1]{em95} do not apply to our $r_o$ defined as a radius where $\partial T/\partial z = 0$ (which may or may not additionally feature the property $K_2 = 0$).

If, on the other hand, one wants to keep $K_2 = 0$ but still bring Eqs.~(\ref{one}) and (\ref{two}) in agreement, one opportunity is to change Eq.~(\ref{two}) by relaxing the assumption of isothermy $\partial T/\partial r = 0$ at $r = r_{\rm max}$.  However, then the expression for heat input in Eq.~(\ref{wc}) will have to be changed, as heat input
will no longer be isothermal. The problem will arise of how to specify the temperature changes. Relaxing the isothermy assumption $\partial T/\partial z = 0$ at $z = z_o$ in the outflow where $K_2 = 0$, as did \citet{emanuel11}, poses the same problem. \citet{emanuel11} postulated certain dynamic constraints in the outflow layer, but their generality has been questioned \citep{montgomery2019}. In any case, when isothermy is abandoned, at $r = r_{\rm max}$ or $r = r_o$, Carnot efficiency cannot describe the thermodynamic cycle $a-c-o-o'-a$. Note also that if at $z=z_o$ we have $\partial T/\partial z \ne 0$, this means that the air parcels emanating from the region of maximum wind do not reach the tropopause.

A final option is to preserve isothermy and $K_2 = 0$ but relax the saturation assumption. This choice appears least ``costly'' for E-PI formulations, since the expression for heat input in the last equality of Eq.~(\ref{wc}) will not change
(although $\delta Q$ will no longer be equal to $T ds^*$).  For example, if relative humidity increases towards the eyewall as much as the air pressure declines, $-dp/p = d\mathcal{H}/\mathcal{H}$ in Eq.~(\ref{qs}), then $\mathcal{C}_2$ in Eq.~(\ref{two}) will be replaced by $\mathcal{C}_2/2$ and Eqs.~(\ref{one}) and (\ref{two}) can match. Here the problem is that assuming $\partial \mathcal{H}/\partial r < 0$ implies $\mathcal{H} < 1$, such that the cycle is no longer reversible and Carnot efficiency in all formulae should be replaced by a smaller value. Fortunately, for large $\varepsilon_C = (T_a - T_o)/T_a \approx 0.3$ the reduction will not exceed a few per cent \citep[see][his~Eq.~25]{pa11}. However, even when $\partial \mathcal{H} /\partial r \ne 0$ the value of $K_2$ can still be large (this happens if $r_o$ is sufficiently small) and Eq.~(\ref{beL}) will overestimate the observed maximum velocity.

Generally, the horizontal change of relative humidity in the boundary layer is the dominant term in Eq.~(\ref{qs}).
For a typical hurricane with a total pressure drop about 50~hPa of which 10~hPa correspond to the eye,
the relative difference between pressure at the radius of maximum wind and the ambient pressure of 1010~hPa will be around $-\delta p/p \approx 0.04$, which is about one fifth
of what the relative humidity contributes, $\delta \mathcal{H}/\mathcal{H} \approx 0.2$ for $\mathcal{H}_a = 80\%$.
Equations~(\ref{one}) and (\ref{two}) indicate that if $d\mathcal{H}/\mathcal{H} \gg -dp/p$, then Eq.~(\ref{two}) with $K_2  =0$ will underestimate
maximum velocity. If $d\mathcal{H}/\mathcal{H} \ll -dp/p$, the opposite will be true. This is a possible, E-PI independent, explanation
for why the majority of tropical cyclones never reach their conventional E-PI $V_{\rm max}$ (while some exceed it). Testing this proposition
requires information on the radial gradients of relative humidity, pressure and temperature, as per Eq.~(\ref{one}), at the radius of maximum wind.

\section{Scaling of maximum velocity}
\label{scale}

The E-PI concept is credited for being formulated in terms of  ``external conditions'', which permits the prediction of maximum storm velocity from environmental parameters \citep[e.g.,][]{bryan09b}. However, these environmental parameters are not known before the storm develops -- they arise {\it within and during} the storm at the radius of maximum wind. Additional assumptions are needed for E-PI to forecast maximum potential intensity.

The magnitude of $\Delta q$ in Eq.~(\ref{vmax1}), as a measure of thermodynamic disequilibrium between the atmosphere and the ocean, cannot be retrieved from consideration of the Carnot cycle.  In E-PI it is assumed that
relative humidity $\mathcal{H}$ of surface air at the radius of maximum wind is equal to its ambient value
$\mathcal{H}_{a}$ \citep[e.g.,][p.~3971]{em95}:
\begin{equation} \label{dq}
\Delta q = (1 - \mathcal{H}_a)q_{sc}^*,
\end{equation}
where low index $c$ specifies that the saturated mixing ratio $q_s^*$, corrresponding to sea surface temperature, is evaluated at radius $r_c$ of maximum wind.

The difference between the saturated mixing ratio ($\mathcal{H}_c=100\%$) at the radius of maximum wind and
the ambient mixing ratio ($\mathcal{H}_a < 100\%$)  along an isothermal surface where the saturated pressure of the water vapor is constant
(and hence the second term in Eq.~(\ref{qs}) is zero)  is a sum of the relative differences of pressure and humidity, Eq.~(\ref{qs}). Neglecting the pressure contribution, which, as we discussed in Section~\ref{impl}, is usually minor as compared to the humidity contribution, the maximum velocity scale in E-PI depends on the water vapor disequilibrium $\Delta q_a$ at the surface in the ambient environment,
$\Delta q \approx \Delta q_a = (1 - \mathcal{H}_a)q_{sa}^*$ \citep[e.g.,][Eq.~38]{em89}:
\begin{equation}\label{sc}
V_{\rm max}^2 \sim \frac{\varepsilon_C}{1-\varepsilon_C} \frac{C_k}{C_D} L (1 - \mathcal{H}_a) q_{sa}^*.
\end{equation}
Compared to Eq.~(\ref{vmax1}), Eq.~(\ref{sc}) neglects the gravitational power of precipitation.
Compared to Eq.~38 of \citet{em89}, Eq.~(\ref{sc}) contains factors $1/(1-\varepsilon_C)$ and $C_k/C_D$.
Using the ideal gas law and our Eq.~(\ref{qdef}) we can re-write Eq.~(\ref{sc}) as
\begin{equation}\label{sc1}
V_{\rm max}^2 \sim \left(\frac{1}{2}\frac{\varepsilon_C}{1-\varepsilon_C} \frac{C_k}{C_D}\frac{\mathcal{L}}{RT_s} \frac{1 - \mathcal{H}_a}{\mathcal{H}_a}\right)
\frac{2 p_{vsa}}{\rho_{sa}},
\end{equation}
where $p_{vsa} = \mathcal{H}_ap_{vsa}^*$ is the actual partial pressure of water vapor in surface air in the ambient environment, $p_{vsa}^*$ is the saturated partial pressure of water vapor at sea surface temperature in the ambient environment, $\rho_{sa}$ is ambient air density at the surface. Using typical tropical values $T_s = 300$~K, $\mathcal{H}_a = 0.8$, $\rho_{sa} \approx 1.2$~kg~m$^{-3}$,  $\varepsilon_C = 0.32$ \citep[see, e.g., Table~1 of][]{em89} and $C_k/C_D = 1$, we have $p_{vs} = 28$~hPa and $V_{\rm max} = 70$~m~s$^{-1}$. Neglecting the factor $1/(1-\varepsilon_C)$ (as it is neglected in \citet{em89}'s Eq.~38 and \citet{em97}'s Eq.~8), we obtain $V_{\rm max} \approx 60$~m~s$^{-1}$ in agreement with Table~1 of \citet{em89}.

The coefficient in parentheses in Eq.~(\ref{sc1}) for the same typical parameters is close to unity:
\begin{equation}\label{fact}
\frac{1}{2}\frac{\varepsilon_C}{1-\varepsilon_C} \frac{C_k}{C_D} \frac{\mathcal{L}}{RT_s} \frac{1 - \mathcal{H}_a}{\mathcal{H}_a} \approx 1.
\end{equation}
This means that numerically the scaling of maximum velocity in E-PI practically coincides with the scaling
\begin{equation}\label{osc}
\rho \frac{V_{\rm max}^2}{2} = p_{vsa}
\end{equation}
proposed within the concept of condensation-induced atmospheric dynamics \citep{pla09,pla11,acp13,ar17}.
The exponential dependence on surface temperature is captured
within the term $p_{vs}$ that is common to both (\ref{sc1}) and (\ref{osc}).
Unlike $p_{vs}$, the factor (\ref{fact}) of E-PI depends only weakly on $T_s$.
(More specifically, the maximum velocity scale in condensation-induced dynamics is $\beta p_{vsa}$,
where $\beta \sim 0.5$ is the degree of water vapor condensation in the outflow \citep{pla11}.
But likewise a more appropriate value for $C_k/C_D$ has been suggested
to be $0.5$ \citep{bell12}. So if both factors are taken into account,
the agreement between the two approaches persists, with both yielding $\rho V_{\rm max}^2/2 \sim 0.5 p_{vsa}$.)

While under typical tropical conditions Eqs.~(\ref{sc1}) and (\ref{osc}) are numerically similar, their physical interpretations differ. In condensation-induced dynamics the partial pressure
of water vapor is interpreted as the local store of available potential energy
which can be converted to the kinetic energy of winds upon condensation. If $\mathcal{H} = 1$,
$p_{vsa}$ and hence the storm intensity are maximum. In comparison, $(1- \mathcal{H})$ in E-PI is a measure of the
disequilibrium between the atmosphere and the ocean. If $\mathcal{H} = 1$, the storm does not develop.
Condensation-induced dynamics explains that in the Earth's gravitational field $p_{vsa}$ is a measure
of dynamic disequilibrium, since the air cannot rise adiabatically without water vapor changing state and impacting pressure gradients. In comparison, a question faced by E-PI as a theory is the nature of the thermodynamic disequilibrium
between the atmosphere and the ocean. How is this disequilibrium maintained and what determines its magnitude?

Intense evaporation in the region of maximum winds should have rapidly driven relative humidity to unity.
Observed relative humidities in intense storms are indeed close to 100\%, conspicuous examples include hurricanes Isabel
\citep[][Fig.~4d]{montgomery06} and Earl \citep[][Fig.~9d]{jaimes15}. For example, in hurricane Isabel relative humidity rose from 75\% outside the storm to 97\% in the eyewall. With relative humidity close to 100\%, moisture input from the ocean in real hurricanes occurs due to the lower temperature of the surface air.

But why should the air cool, by how much and under which circumstances? Namely these factors, and not the ambient relative humidity as Eq.~(\ref{dq}) presumes, would determine E-PI maximum velocity.
According to observations, in the region of maximum wind there is a temperature difference $\Delta T$ of a few degrees Kelvin between the sea surface and the adjacent air \citep[e.g.,][Fig.~9c]{jaimes15}. For example,
in hurricane Isabel the surface air cooled by 4~K as it moved from the storm's outskirts
towards the radius of maximum wind. (Another reason for air cooling, as suggested by Dr. Kerry Emanuel (pers. comm.) is the re-evaporation of the falling droplets.)  This temperature difference not only accounts for the sensible heat flux but also determines $\Delta q > 0$ which allows for moisture input from the ocean.

However, as we discussed in Section~\ref{dh}, sensible heat has never been theoretically assessed
within the E-PI concept. Some confusion in the literature reflects this situation. For example, according to \citet{garner15} quotes \citet{em86} and states that E-PI assumes a surface air-temperature deficit of 1$^{\rm o}$-2$^{\rm o}$C.
Meanwhile \citet[][Table~1]{holland97} considers that in E-PI the sea surface temperature equals air temperature ($\Delta T = 0$).
Indeed, \citet[][p.~591]{em86} assumed $\Delta T = 0$ in his first comparison of E-PI with empirical data.
(Note that when $\Delta T = 0$ and $J_S = 0$, \citet{bister98}'s formula, Eq.~(\ref{be}), coincides with our revised estimate, Eq.~(\ref{beL}).)

We conclude that the surface moisture flux that governs $V_{\rm max}$ in E-PI is itself governed by parameter $\Delta T$. Yet the E-PI concept does not make any predictions or justifiable assumptions  about $\Delta T$. How can Eq.~(\ref{dq}) be invalid for describing observed $\Delta q$ and still produce a valid $V_{\rm max}$ in Eq.~(\ref{sc})?
Equation~(\ref{osc}) explains that this is because the E-PI parameters combine into $p_{vsa}$. If condensation-induced dynamics determines $V_{\rm max}$ from $p_{vsa}$, so will any other formulation, irrespective of whether its underlying physical concepts are valid, if that formulation numerically coincides with Eq.~(\ref{osc}). This happens in Eq.~(\ref{sc}) of E-PI.

\section{Discussion}
\label{disc}

The E-PI concept comprises three interlocking blocks. The first block is the relationship between the volume-specific rates of heat input, work, turbulent dissipation and kinetic energy generation at the top of the boundary layer, Eq.~(\ref{pow}).
This block comprises several dynamic and thermodynamic assumptions about the atmosphere at the top of, and above,
the boundary layer: in brief, the air is isothermal and saturated at $z = h_b$ and the air motion is inviscid and adiabatic at $z > h_b$. The second block pertains to the atmosphere within the boundary layer: it is the proportionality
between the volume-specific rates of heat input and turbulent dissipation at $z = h_b$ and their surface-specific fluxes at $z = 0$, Eq.~(\ref{dj}). The two blocks relate maximum velocity to the degree of water vapor disequilibrium $\Delta q$
between the atmosphere and the ocean at the radius of maximum wind, Eq.~(\ref{beL}). This disequilibrium is {\it a priori}
unknown requiring a third block which consists of an assumption about how this local disequilibrium relates to environmental
parameters outside the storm (Section~\ref{scale}). This final block is not required for testing E-PI in models,
but is crucial for the link between E-PI and observations. It is also the least analyzed aspect of E-PI.

Here we re-derived the E-PI concept separating dynamics from thermodynamics, paying special attention to assumptions
and generalizing where possible. The starting point of the original E-PI concept was to assume
that surfaces of constant angular momentum are also surfaces of constant moist entropy \citep{em86}.
We showed that it is possible to derive E-PI without considering entropy.
We lifted the assumption of gradient wind balance and showed that the E-PI formulation for a radially unbalanced inviscid atmosphere does not change, cf. Eqs.~(\ref{vmax}) and (\ref{V}).
We also showed that, contrary to previous research, the energy
expended to lift water has little impact on storm intensity (Section~\ref{sHP}).

Neglecting this energy, the logic of E-PI's first block can be summarized as follows. First, the total work $A^+ + A^-$ of the imaginary steady-state cycle, Fig.~\ref{fig1} and Eq.~(\ref{oint}), is equal to $\varepsilon_C$ times the isothermal heat input from the ocean, Eq.~(\ref{wc}). Second, from the first law of thermodynamics this isothermal heat input
is the sum of $A^+$ (work in the boundary layer) and latent heat input. Third, from the Bernoulli equation
work $A^+$ on a horizontal streamline is equal to the radial increment of kinetic energy minus turbulent friction losses, Eq.~(\ref{fdl}). Forth, E-PI assumes that $A^-$ (work above the boundary layer) is negative and (approximately)
equal in magnitude to the positive radial increment of kinetic energy in the boundary layer, Eq.~(\ref{wup}).
This means that working against the pressure gradient in the upper atmosphere
consumes all kinetic energy the air has acquired in the boundary layer.

These four statements relate the radial increment of kinetic energy, turbulent friction losses
and latent heat input. By definition, at the radius of maximum wind the radial
increment of kinetic energy is zero. Thus at the radius of maximum wind E-PI yields a relationship
between local latent heat input and turbulent friction, Eq.~(\ref{djl}),
from which E-PI maximum velocity ultimately derives, Eq.~(\ref{vmax1}).
These relationships are formulated in the differential form for an infinitely narrow cycle enclosing the streamline
that emanates from the boundary layer at the radius of maximum wind $r = r_{\rm max}$.

Thermodynamics enters explicitly only as parameter $\varepsilon_C$, which applies
when the considered cycle is a Carnot cycle. If it is not, then a lower value $\varepsilon < \varepsilon_C$ should
be used. Implicitly, for the Carnot formulation to be applicable, several assumptions must
be fulfilled along the streamline, including isothermy $\partial T/\partial r = 0$ at the radius of maximum wind
and $\partial T/\partial z = 0$ in the outflow at $r = r_o$. The streamline itself must be an adiabat.

The above logic works when the kinetic energy increment in the outflow at $r = r_o$ is negligible compared to pressure work at the top of the boundary layer, $K_2 = 0$ in Eq.~(\ref{pow}). In the original formulation of E-PI it was assumed that the streamlines ascending from the boundary layer in the region of maximum winds
reach the top of the troposphere where $\partial T/\partial z = 0$ at a large radius $r_o \gg r_{\rm max}$. In this case the contribution of the outflow to the storm's power budget was shown to be negligible
unless $r_o$ is very large \citep[e.g.,][their Eq.~34]{bister98}, because air velocity rapidly declines with
distance from the storm center in the cyclonic part of the storm.  Here we showed that these assumptions are not
compatible with other E-PI assumptions (Section~\ref{cons}).
If isothermy is reached at a small radius $r_o$ comparable to $r_{\rm max}$, see Eq.~(\ref{rmro}), then $K_2$
is large and significantly impacts E-PI formulations. We showed that in such a case discarding $K_2$,
as the conventional E-PI formula (\ref{vmax}) and all its modifications, (\ref{V}), (\ref{be}), (\ref{beL}) do,
will result in E-PI significantly underpredicting maximum velocity.

We re-analyzed the relationship
between the oceanic sensible heat flux and dissipative heating, which resulted in a revised formula for E-PI velocity
depending on latent heat flux only, Eq.~(\ref{beL}).
We showed that the formula of \citet{bister98}, intended to account for dissipative heating,
overestimates E-PI intensity due to the (unjustifiably) neglected compensation between
external and dissipative heating (Section~\ref{dh}).
To our knowledge, these relationships have have not been previously described. Likely this reflects that the flux of heat has not previously been formulated in terms of the radial gradients of $p$ and $q$. Previously all derivations considered potential temperature $\theta$ and entropy, which masked some of the relevant physical relationships.
In this formulism, sensible heat input $Q_S$ for an isothermal process is written not as
$Q_S = -\int \alpha_d dp$ but in terms of entropy and the Gibbs function \citep[e.g.,][p.~96]{pa11},
which obscures the equivalence between work and heat input. This equivalence matters,
because at the radius of maximum wind, to which E-PI pertains, rates of work and turbulent dissipation
are equal, Eq.~(\ref{turb}). This means that if the dissipative heating grows,
the external (oceanic) flux of sensible heat must diminish, and vice versa.
Another reason why these effects and Eq.~(\ref{vmax1}) were not previously formulated is
because E-PI did not use $\partial V/\partial r = 0$ (the condition for maximum velocity) to derive Eq.~(\ref{vmax})
\citep[a point noted by][]{montgomery17b}.

Unlike numerical models of tropical storms tuned to produce the desired patterns,
the theoretical E-PI concept seeks to explain why real hurricanes typically have a maximum speed of approximately 60~m~s$^{-1}$ on the basis of verifiable assumptions. The main result of E-PI, the culmination of all theorizing within the first two blocks, is to show that the storm intensity depends on the thermodynamic disequilibrium between
the atmosphere and the ocean and describe the peculiarities of this dependence.

We emphasize that once the nature and magnitude of a (thermo)dynamic disequilibrium is given, it is possible to spin up a model storm of practically any intensity by varying the turbulence parameters. But if the disequilibrium is physically unrealistic, such models, including those tuned to produce observable winds, will be of limited help in understanding and predicting intensity of {\it real} storms. For example, \citet{mrowiec11} described ``dry hurricanes'' that are driven by the sensible heat flux from the ocean in a dry atmosphere. In their model
the thermal disequilibrium between the air and the ocean is an arbitrary choice for the modeller.
It is artificially maintained by setting a prescribed temperature difference $\Delta T$ between the air and the ocean.
(Note that due to the interrelationship between the sensible heat flux and pressure work, Eq.~(\ref{jls}),
once $\Delta T$ is prescribed, a radial pressure gradient is guaranteed such models.)
However, since a key issue -- why the thermal disequilibrium persists in spite of the thermal flux that destroys it --
cannot be addressed in these models, their realism and relevance remain undetermined.

Most theoretical efforts within E-PI are devoted to the first and second block, i.e. to the link between the properties of air
circulation and the surface heat fluxes. What determines the magnitude of the disequilibrium and how it is maintained
remains unresolved and little discussed in the literature. Unlike the ``dry models'' where the disequilibrium is entirely arbitrary, E-PI  does link this disequilibrium to an important environmental parameter: the undersaturation of the ambient boundary layer, $\mathcal{H}_a < 100\%$, Eq.~(\ref{dq}). We showed, however, that this assumption does not match the observations: the relative humidity in the region of maximum wind is close to 100\%. Then  the moisture flux from the ocean is governed by the temperature difference $\Delta T$, for which
E-PI does not make any specifications.

We have shown that the quantitative agreement of E-PI velocity scale with observations, Eq.~(\ref{sc}),
may have a distinct explanation consistent with the concept of condensation-induced atmospheric dynamics.
We gave an alternative explanation to the maximum velocity scaling having
shown that maximum E-PI kinetic energy is equal to the partial pressure of water vapor at the surface,
Eq.~(\ref{osc}). This scaling is central to the concept of condensation-induced atmospheric dynamics \citep{pla14,pla15}.
In the Earth's gravitational field partial pressure of water vapor is a measure of dynamic disequilibrium.
For condensation-induced hurricanes, the key process is the positive feedback between the radial air motion and the pressure drop at the surface associated with condensation and hydrostatic adjustment. As air streams towards the hurricane
center and ascends, the water vapor condenses and the air pressure drops as determined by the partial pressure
of water vapor at the surface. The disequilibrium persists until the atmosphere dries out by precipitation.
(Note that the fact that one can simulate dry hurricanes
in a mathematical model does not mean that real hurricanes are not driven by condensation dynamics.)

This concept requires more attention and we value any feedback. One reviewer
noted that condensation cannot lead to considerable pressure gradients because
droplets that form upon condensation are falling with terminal velocity and thus their weight
compensate for all possible pressure drop due to water vapor removal. This would make sense,
at least in the vertical dimension, if all condensed moisture remained in the air.
However, as discussed by \citet{jgra17},
at the moment a droplet forms it has the same velocity as the air and does not impose any velocity-related force on it.
The droplets accelerate relative to the air, and when they reach their terminal velocity, this velocity is  so large that
most of condensed moisture is removed from the air while it is ascending. Even in hurricanes
the amount of condensed moisture remaining in moist air is about one percent of the original water vapor.
Thus this residual condensed moisture cannot compensate
for the condensation-induced pressure perturbations.

Dr. Chanh Kieu (pers. comm.) noted that if, in a model hurricane, surface fluxes are switched off,
the storm does not intensify. This does not conflict with our understanding of condensation-induced dynamics,
in which storms require moist air to persist. In current models of motionless hurricanes
the only source of moisture is the ocean, so if this flux discontinues the atmosphere dries
and any moisture-driven would necessarily cease. But in real storms most moisture
derives not from concurrent evaporation but from previously
accumulated water vapor   in the atmospheric air that feed into the system \citep{ar17}.
The motionless model storms lack access to such moisture and depend solely (and artificially) on the ocean.

At the same time, we believe that surface heat fluxes remain relevant to our broader dynamic interpretation of E-PI, that
the increment of kinetic energy from in the boundary layer must be sufficient for the air to overcome the negative pressure gradient in the upper atmosphere, Eq.~(\ref{oint}). The air must have sufficient energy to flow away from the hurricane. If not generated in the boundary layer, this energy could derive from a pressure gradient in the upper atmosphere:  if, at the expense of the hurricane's extra warmth,  the air pressure in the column above the area of maximum wind is higher than in the ambient environment, this pressure gradient will accelerate the air outward. However, a significant pressure deficit at the surface precludes the formation of a significant pressure surplus aloft \citep[e.g.,][Fig.~1d]{tellus17}.

Moreover, this pressure deficit is what accelerates air in the boundary layer. If the pressure gradient is sufficiently steep and the radial motion sufficiently rapid, the expansion of air will be accompanied by a drop of temperature (i.e. the process will be closer to an adiabat than to an isotherm). In Hurricane Isabel the surface air cooled by about 4~K between the outer core (150-250~km) and the eyewall (40-50~km) \citep[][Fig.~4c]{montgomery06}, while pressure fell from less than 1013~hPa to approx 960~hPa at the eyewall \citep[][Fig.~4]{aberson06}. (Air pressure
at the outermost closed isobar $\sim 465$~km from the center was 1013~hPa, hence at 150-250~km from the center it should have been smaller.) This is almost a dry adiabatic process with $dp/p = (C_p/R) dT/T$, where $C_p = (7/2) R$ is molar heat capacity of air at constant pressure. (Likewise Eq.~(\ref{qs}) gives $dq/q \approx 0$ with the observed change of relative humidity from 80\% to 97\% over the same distance.) The surface air streams towards the center so rapidly that it lacks time to take much heat from the ocean.

If the warm air creates a pressure surplus aloft facilitating the outflow, cold air, conversely, creates a pressure deficit. This enhances the pressure gradient in the upper atmosphere against which the air must work to leave the hurricane. Consequently, the storm cannot deepen indefinitely. Eventually {\it the kinetic energy acquired at the surface} becomes insufficient  for the rising and adiabatically cooling air to overcome the pressure gradient in the upper atmosphere, and the outflow must weaken. This condition provides distinct constraints on storm intensity. Further research is needed to see
whether such processes are relevant in real storms.

%\appendix
\setcounter{section}{0}%
\setcounter{equation}{0}%
\renewcommand{\theequation}{A.\arabic{equation}}%

\section*{\begin{center}  Appendix A: Equivalence between our Eq. (\ref{f18}) and Eq.~18 of \citet{em86} \end{center} }
\label{A1}
\citet{em86} assumed that the air is in gradient wind balance above the boundary layer:
\begin{equation}\label{gwb}
\alpha \frac{\partial p}{\partial r} = \frac{v^2}{r} + fv,  \quad  z \ge h_b.
\end{equation}
Here $v$ is tangential velocity; under approximations (\ref{he}) and (\ref{gwb}) it is equal to total velocity (the radial and vertical velocities are negligible). The Coriolis parameter $f \equiv 2 \Omega \sin \varphi$ is assumed constant ($\varphi$ is latitude, $\Omega$ is the angular velocity of Earth's rotation).

In hydrostatic equilibrim (\ref{he}) for any closed contour in the atmosphere
\begin{equation}\label{cl}
\oint \alpha dp = \oint \alpha \frac{\partial p}{\partial r} dr + \oint \alpha \frac{\partial p}{\partial z} dz = \oint \alpha \frac{\partial p}{\partial r} dr.
\end{equation}

Using (\ref{gwb}) for any path $x-y$ along which the angular momentum
\begin{equation}\label{M}
M \equiv vr + \frac{f r^2}{2}
\end{equation}
\noindent
is constant, one obtains
\begin{equation}\label{co}
-\int_x^y \alpha \frac{\partial p}{\partial r} dr = \frac{v^2}{2}\bigg|_x^y.
\end{equation}

Equation~(\ref{co}) can be obtained directly from the Bernoulli equation by using hydrostatic equilibrium (\ref{he}) and putting $V = v$ and $\mathbf{f} = 0$ in Eq.~(\ref{B}). Another way to obtain (\ref{co}) is from gradient wind balance equation (\ref{gwb}), which using Eq.~(\ref{M}) and
\begin{equation}\label{v}
v = \frac{M}{r} - \frac{fr}{2},
\end{equation}
we can write as
\begin{equation}\label{gwb1}
\alpha \frac{\partial p}{\partial r} = v(v+fr)\frac{1}{r} = \left(\frac{M}{r} - \frac{fr}{2}\right)\left(\frac{M}{r} + \frac{fr}{2}\right)\frac{1}{r}
= \left(\frac{M^2}{r^2} - \frac{f^2r^2}{4}\right)\frac{1}{r}.
\end{equation}
For a trajectory $x-y$ along which $M$ and $f$ are constant, we obtain Eq.~(\ref{co}) from Eq.~(\ref{gwb1})
%\begin{linenomath*}
\begin{multline}\label{vg}
-\int_x^y \alpha \frac{\partial p}{\partial r} dr = -\int_x^y \left(\frac{M^2}{r^2} - \frac{f^2r^2}{4}\right)\frac{dr}{r}
= \frac{1}{2}\left(\frac{M^2}{r^2} + \frac{f^2r^2}{4}\right)\bigg|_x^y  \\
\equiv \frac{1}{2}\left(\frac{M^2}{r^2} -Mf + \frac{f^2r^2}{4}\right)\bigg|_x^y \equiv
\frac{1}{2}\left(\frac{M}{r} - \frac{fr}{2} \right)^2\bigg|_x^y \equiv \frac{v^2}{2}\bigg|_x^y.
\end{multline}
%\end{linenomath*}

From Eqs.~(\ref{he}), (\ref{cl}) and (\ref{co}) we obtain
\begin{equation}\label{aco}
-\oint \alpha dp = -\int_a^c \alpha \frac{\partial p}{\partial r} dr
 - \frac{v_c^2 - v_a^2}{2} -  \frac{v_{o'}^2 - v_o^2}{2},
\end{equation}
\noindent
which is equivalent to Eq.~(\ref{oint}) where $V=v$. Combining Eq.~(\ref{aco}) with Eq.~(\ref{wc})
we obtain Eq.~(\ref{f18}), where $V$ is replaced with $v$.

\citet{em86} assumed that air streamlines $c - o$ and $a - o'$ conserve angular momentum:
\begin{equation}\label{Mo}
M_a = M_{o'},\quad M_c = M_o.
\end{equation}

Taking into account that, according to (\ref{gwb}) and (\ref{M}),
\begin{equation}\label{V2}
v^2 = r \alpha \frac{\partial p}{\partial r} - fvr = r \alpha \frac{\partial p}{\partial r}  - f M + \frac{f^2r^2}{2},
\end{equation}
using Eq.~(\ref{Mo}), recalling that $r_o = r_{o'}$ and assuming, following \citet{em86},
that for  $r_a  \le r \le r_o$ the radial pressure gradient is sufficiently small for the first term in the right-hand side of (\ref{V2})
to be neglected, we obtain from Eq.~(\ref{aco})
\begin{equation}\label{tw}
-\oint \alpha dp = -\int_a^c \alpha \frac{\partial p}{\partial r} dr - \frac{1}{2}  r \alpha \frac{\partial p}{\partial r} \Big|_c - \frac{f^2}{4}\left(r_c^2 - r_a^2\right).
\end{equation}

In the case of E-PI the cycle's efficiency is equal to Carnot efficiency, $\varepsilon_C = (T_a - T_o)/T_a$. Combining
(\ref{aco}) and (\ref{wc}) and using (\ref{tw}) we find
\begin{equation}\label{f18b}
\varepsilon_C \int_a^c \left(-\alpha_d \frac{\partial p}{\partial r} + L\frac{\partial q}{\partial r}\right)dr = -\int_a^c \alpha \frac{\partial p}{\partial r} dr
- \frac{1}{2} r \alpha \frac{\partial p}{\partial r} \Big|_c - \frac{f^2}{4}\left(r_c^2 - r_a^2\right) - \oint \alpha q dp.
\end{equation}

Using the definition of equivalent potential temperature $\theta_e=\theta \exp \left(L q/c_p T \right)$,  the Exner function $\pi\equiv  (p/p_0)^{R/C_p}
=T/\theta$ \citep[][Eq.~15]{em86}, where $\theta$ is the potential temperature and $R/C_p=2/7$, assuming $\alpha \approx \alpha_d$ is constant, changing the notations $r_c \to r$, $r_a \to r_0$, $T_a \to T_B$, $T_o \to \overline{T}_{\rm out}$ and neglecting the last term in the right-hand side of (\ref{f18b}), we obtain Eq.~18 of \citet{em86} from Eq.~(\ref{f18b}):
\begin{equation}
-\frac{T_B - \overline{T}_{\rm out}}{T_B} \ln \frac{\theta_e}{\theta_{ea}} = \ln \frac{\pi}{\pi_a}
+ \frac{1}{2} r \frac{\partial \ln \pi}{\partial r} + \frac{1}{4}\frac{f^2}{c_p T_B}\left(r^2 - r_0^2\right).
\end{equation}

\setcounter{section}{0}%
\setcounter{equation}{0}%
\renewcommand{\theequation}{B.\arabic{equation}}%
\section*{\begin{center} Appendix B: Deriving Eq.~(\ref{est}) \end{center} }
\label{A2}
%\begin{linenomath*}
%\begin{multline}  \label{est1}
\begin{equation}\label{est1}
- \oint \alpha q dp = - \oint \alpha q \frac{\partial p}{\partial r} dr + \oint q g dz = %\\
- \int_a^c \alpha q \frac{\partial p}{\partial r}dr - \int_c^o \alpha q \frac{\partial p}{\partial r} dr
- \int_{o'}^a \alpha q \frac{\partial p}{\partial r} dr - \oint gz \frac{\partial q}{\partial z} dz  
\end{equation}
%\end{multline}
%\end{linenomath*}
%\begin{linenomath*}
%\begin{multline} \label{est2}
\begin{equation}  \label{est2}
= - \int_a^c \alpha q \frac{\partial p}{\partial r}dr
- \frac{1}{2}\left(q_cV_c^2 - q_aV_a^2 + q_{o'}V_{o'}^2 - q_oV_o^2\right) % \\
- \int_c^o \frac{V^2}{2} \frac{\partial q}{\partial r} dr
- \int_{o'}^a \frac{V^2}{2}\frac{\partial q}{\partial r} dr - \oint gz \frac{\partial q}{\partial z} dz
\end{equation}
%\end{multline}
%\end{linenomath*}
%\begin{linenomath*}
%\begin{multline}\label{est3}
\begin{equation}\label{est3}
= -\int_a^c q\left( \alpha \frac{\partial p}{\partial r} +\frac{1}{2}\frac{\partial V^2}{\partial r} \right) dr
- \frac{1}{2}\left(q_{o'}V_{o'}^2 - q_oV_o^2\right)
- \oint \frac{V^2}{2} \frac{\partial q}{\partial r} dr
- \oint gz \frac{\partial q}{\partial z} dz
\end{equation}
%\end{multline}
%\end{linenomath*}

\setcounter{section}{0}%
\setcounter{equation}{0}%
\renewcommand{\theequation}{C.\arabic{equation}}%
\section*{\begin{center} Appendix C: Estimating $K_2$ in Eq.~(\ref{alg}) \end{center} }
\label{A3}
We assume following \citet{em86} that the atmosphere is in gradient wind balance (\ref{gwb})
and that streamlines $a - o'$ and $c-o$ conserve angular momentum (\ref{M}).
From Eqs.~(\ref{M}) and ~(\ref{Mo}) we have
\begin{equation}\label{Vo2}
M_a - M_c = v_a r_a - v_c r_c + \frac{f}{2}(r_a^2 -r_c^2) = (v_{o'} - v_o) r_o = M_{o'} - M_o
\end{equation}
and
\begin{equation}\label{Vo3}
v_a (r_a - r_c) + (v_a - v_c) r_c + \frac{f}{2} (r_a - r_c)(r_a + r_c) = (v_{o'} - v_o) r_o.
\end{equation}
Dividing Eq.~(\ref{Vo3}) by $r_o (r_a - r_c)$ we obtain
\begin{equation}\label{Vo4}
\frac{1}{r_o}\left[ v_a + \frac{v_a - v_c}{r_a - r_c} r_c + \frac{f}{2} (r_a + r_c) \right] = \frac{v_{o'} - v_o}{r_a - r_c}.
\end{equation}
In the limit $r_a \to r_c$ we find from Eq.~(\ref{Vo4})
\begin{equation}\label{Vo5}
\frac {\partial v_o}{\partial r} = \frac{r}{r_o} \frac{\partial v}{\partial r} + \frac{v}{r_o} + f \frac{r}{r_o},\,\,\,\,\,z = h_b.
\end{equation}

To obtain the same result in a more straightforward manner,
using Eqs.~(\ref{M}) and ~(\ref{Mo}) we can express $v_o$ as
\begin{equation}\label{voo}
v_o = \frac{1}{r_o} \left[vr + \frac{f}{2} (r^2 - r_o^2)\right],
\end{equation}
where $v = v_c$ and $r = r_c$.
Taking the derivative of the right-hand side of Eq.~(\ref{voo}) over $r$ with $\partial r_o /\partial r = 0$
we find
\begin{equation}\label{voo2}
\frac{1}{2}\frac{\partial v_o^2}{\partial r} = \frac{1}{r_o^2} \left[vr + \frac{f}{2} (r^2 - r_o^2)\right]
\left(r \frac{\partial v}{\partial r} + v + f r \right),  \quad  z = h_b.
\end{equation}
At the radius $r = r_{\rm max}$ of maximum wind $\partial v/\partial r = 0$, so using
the gradient wind balance (\ref{gwb}) and neglecting $q_o \ll q \ll 1$ in the definition of $K_2$ (\ref{K}) we have from Eq.~(\ref{voo2})
\begin{equation}\label{K2}
K_2 \equiv \frac{1}{2}\frac{\partial v_o^2}{\partial r} = \frac{v (v + fr)}{r} \left[\frac{r^2}{r_o^2} + \frac{fr}{2v}\left(\frac{r^2}{r_o^2}-1\right)\right] =
\alpha \frac{\partial p}{\partial r} \left[\frac{r^2}{r_o^2} + \frac{fr}{2v}\left(\frac{r^2}{r_o^2}-1\right)\right],
\end{equation}
where $r = r_{\rm max}$, $v = v_{\rm max}$, and $z = h_b$.
The nature and magnitude of $K_2$ in real storms remains to be investigated.
As noted by \citet{smith14}, a mechanism that would provide an increment of angular momentum,
and hence a kinetic energy increment, along $o-o'$ does not appear to exist.
\citet{ar17} indicated that extra angular momentum can arise in the upper atmosphere as
a real steady-state hurricane is an open system that moves through the atmosphere and can import
angular momentum as it does air and water vapor.

%%%%%%%%%%%%%%%%%%%%%%%%%%%%%%%%%%%%%%%%%%%%%%%%%%%%%%%%%%%%%%%%%%%%%
% ACKNOWLEDGMENTS
%%%%%%%%%%%%%%%%%%%%%%%%%%%%%%%%%%%%%%%%%%%%%%%%%%%%%%%%%%%%%%%%%%%%%
%
\acknowledgements

This work is par\-ti\-al\-ly supported by  the University of California Agricultural Experiment Station and the
CNPq/CT-Hidro - GeoClima project Grant~404158/2013-7. We thank Pinaki Chakraborty and Kerry Emanuel for useful comments.
A. Makarieva is grateful to Chanh Kieu for an earlier discussion of atmospheric heat engines. There are no data utilized in this study.

\bibliographystyle{copernicus}
%\bibliography{met-refs}

\end{document}